# Natural reward as the fundamental macroevolutionary force

Owen M. Gilbert[1]

[1]Department of Integrative Biology, The University of Texas at Austin

owen.gilbert@gmail.com (comments welcome)

**Darwin's theory of evolution by natural selection does not predict long-term progress or advancement, nor does it provide a useful way to define or understand these concepts. Nevertheless, the history of life is marked by major trends that appear progressive, and seemingly more advanced forms of life have appeared. To reconcile theory and fact, evolutionists have proposed novel theories that extend natural selection to levels and time frames not justified by the original structure of Darwin's theory. To extend evolutionary theory without violating the most basic tenets of Darwinism, I here identify a separate struggle and an alternative evolutionary force. Owing to the abundant free energy in our universe, there is a struggle for supremacy that naturally rewards those that are first to invent novelties that allow exploitation of untapped resources. This natural reward comes in form of a temporary monopoly, which is granted to those who win a competitive race to innovate. By analogy to human economies, natural selection plays the role of nature's inventor, gradually fashioning inventions to the situation at hand, while natural reward plays the role of nature's entrepreneur, choosing which inventions to first disseminate to large markets. Natural reward leads to progress through a process of invention-conquest macroevolution, in which the dual forces of natural selection and natural reward create and disseminate major innovations. Over vast time frames, natural reward drives the advancement of life by a process of extinction-replacement megaevolution that releases constraints on progress and increases the innovativeness of life.**

*KEY WORDS: invention, entrepreneurship, innovation, success, progress, advancement*

## Introduction

Under Darwin's theory of evolution by natural selection, all beauty and wonder in nature comes from the diversity of life, ultimately created by a single evolutionary force. Darwin captured his sense of beauty with a powerful metaphor. Imagining himself sitting on the edge of a tangled bank, Darwin contemplated how a multitude of organic forms could have been derived from simple laws acting around him: growth with reproduction, inheritance, variability, and a struggle for existence leading to natural selection [1]. Within these laws, Darwin invoked natural selection as the only deterministic force (*deterministic* because it is both non-random and assumes the causes of events come before rather than after events). According to Darwin, natural selection causes slight modifications of form, function, and instinct that adapt a species to its immediate environment. Over time, evolution by natural selection leads to the origin of new complex structures and instincts, divergence of character, and the gradual change of species.

     On the historical backdrop of the dominant scientific debates of the time, the theory of natural selection created a dilemma. On the one hand, it suggested that all species had ultimately descended from one or a few forms. This suggested that major progressive patterns of evolution, as assumed by Lamarck [2] but contended by Lyell [3], were a real aspect of the history of life [4], pp. 337, 356. On the other hand, it did not explain long-term trends of increasing complexity or diversity. In recognition of this, Lyell wrote to Darwin and encouraged him to allow the possibility that natural selection was not the only guiding force and to "modestly limit the pretensions of selection" [5]. Darwin replied by arguing that natural selection was the only force and that Lyell invoked "miraculous additions" [6, 7]. To Darwin's objections, Lyell responded, "I care not for Creation, but I want something higher than Selection" [8].

## The limits of Darwinism

Darwin's first postulate was that the tendency of all organisms to increase in numbers will inevitably outstrip resources [1], pp. 4, 63. With resource limitation, there is a competition between types for limited resources or what Darwin called *the struggle for existence*. Darwin reasoned that the most severe form of competition, leading to displacement of one type by another, would occur within species. Thus, Darwin identified the species as the evolving population. Darwin also reasoned that because the types of organisms competing with each other belong to the same species, their variations would be slight, and modifications of form and function gradual. Darwin explained the origin of complex traits with models that conceptually break them into simpler components, and then chart their historical evolution [9–11]. By creating a historical theory for origins, which uses natural selection as a force acting for immediate benefits and through slight and successive modifications, Darwin advanced biology far beyond the teleological theories of Aristotle and Paley, which assumed that complex traits originate for their ultimate apparent functions [11–15]. Some of Darwin's contemporaries thus recognized that Darwin's theory "demolished" [16], or dealt a "death blow", to teleological thinking in biology [17]; [18], p. 330.

     How exactly did Darwinism challenge teleology? According to Darwin's logic, complex traits do not originate for the ultimate design purpose they appear to serve, as would be the case if they were intentionally designed by a Creator who planned every detail of existence [11–13, 15]. Instead, the course of evolution is determined by simple laws and the deterministic force of natural selection. Darwin [19], p. 51 noted that it was less derogatory to the Creator if He had formulated several simple laws, rather than each parasite and predator independently. More importantly for scientific questions, howev-

**Figure 1. Framework of the theory of natural selection.**

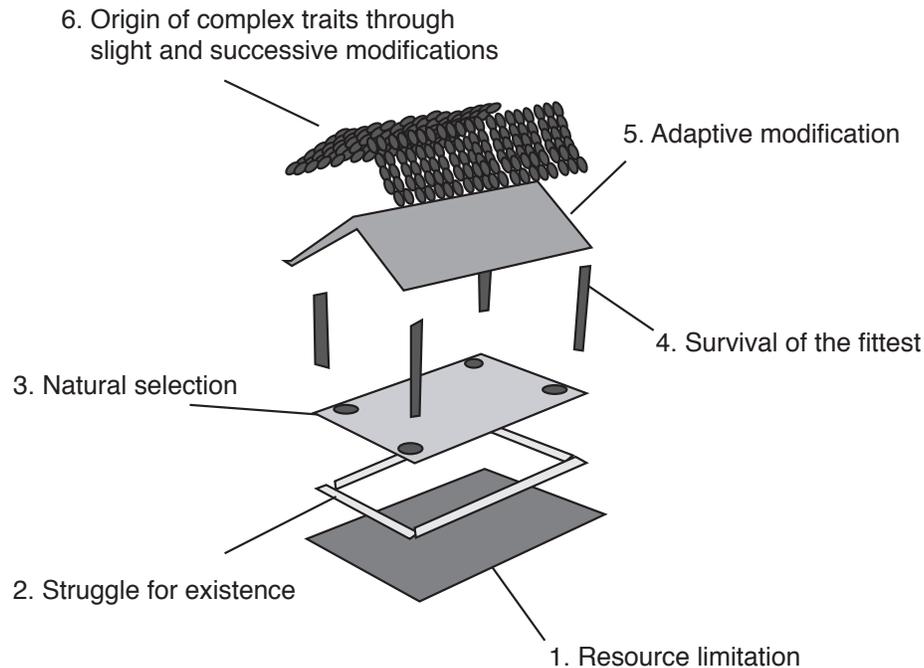

er, Darwin's method provided a way to explain complexity. For traits of moderate complexity, in which subparts originated for a singular purpose (e.g., vision), Darwin's approach was important for drawing attention to the mechanistic ways traits are co-opted from previous functions [9, 10, 20]. For traits of extreme complexity, Darwin's approach showed how subparts can originate for diverse purposes and only incidentally have long-term effects as part of a whole. By breaking complex traits into simpler components and charting historical evolution, Darwin's approach uniquely identifies the causes for origins [9, 11, 21, 22], which may be different from ultimate effects [23–26].

Lacking many detailed examples, Darwin emphasized a principle that would guide historical inquiry. He stressed that natural selection acts only by the accumulation of slight modifications of structure or instinct, each profitable to the individual under its conditions of life [1], pp. 170, 233, 235, 435. Darwin also emphasized that natural selection produces only relative perfection and that true wonder is why imperfection is not more commonly observed [1], p. 472. Darwin [1], p. 133 also stated that natural selection, though sometimes leading to advancement, includes no necessary law of advancement—and could just as easily lead to retrogression [1], p. 137. Darwin thus encouraged an approach that focuses on slight and successive modifications that build complex traits.

Darwin's theory thus advanced biology past former teleological theories [12, 15]. However, it did not answer the prevailing question of the day. In Darwin's time, the great science was geology, and the major question was whether life had progressed with time [27–29]. Darwin's theory required that life had increased in diversity and complexity, but did Darwin's theory explain these apparently progressive trends? To make his theory relevant to progress in general, Darwin argued that species going extinct would tend to have an inferiority held in common [1], pp. 321–322, 327, 344, that natural selection causes newer forms to supplant their less-improved ancestors [1], pp. 5, 108, 119, 126, etc., and that those from newer epochs tend to beat and exterminate those from previous epochs [1], p. 337. Darwin also sometimes suggested that natural selection would necessarily lead to progress towards perfection [1], p. 489. In making these arguments, however, Darwin contradicted himself and Lyell, a long-time critic of naïve progressionism, was quick to point out the logical inconsistencies [5]. As an early harbinger of this situation, Darwin remarked to Hooker [30], "With respect to 'highness' & 'lowness,' my ideas are only eclectic & not very clear".

## How microevolution became "the whole truth"

Almost a century after Darwin first proposed his theory, Dobzhansky [31] began the third edition of his famous *Genetics and the Origin of Species* by stating that, "Simpson's *Tempo and Mode in Evolution* [32] and *Meaning of Evolution* [33] ended the belief…that paleontology has discovered some mysterious 'macroevolution' which is inexplicable in the light of the known principles of genetics". The definition of macroevolution included the origin of highly complex traits, higher taxonomic categories, and the major superspecific trends [34], p. 291. In contrast, microevolution involves changes of demes or species [31, 34]. Dobzhansky [31], p. 17 then went on to identify "the known genetics principles" to include mutation, selection, and genetic drift, and he argued that the works of Simpson and others found nothing in macroevolutionary phenomena that would require other than the known genetic principles for causal explanation. Dobzhansky proposed that the methods of experimental genetics are limited to differences at lower taxonomic levels, and therefore that macroevolutionary phenomena can be understood only as inferred from microevolutionary genetic processes. Following Dobzhansky, authors asked whether microevolutionary genetic processes are sufficient to explain macroevolution, or whether it is necessary to postulate other

"genetic factors" [35] p. 1; [36] p. 969; [37], p. 475. This posing of the question led to a focus on microevolutionary genetics and a distraction away from other deterministic forces. Most modern evolutionists, including those advocating alternative theories, have assumed natural selection is the only deterministic evolutionary force [37–43].

Two ideas reinforced the assumption that macroevolution is simply microevolution continued into vast expanses of time [44], p. 305; [35], p. 1. The first was Dobzhansky's [31] extremely influential [45] description of the adaptive landscape. In Dobzhansky's [31] hands, natural selection became a force that optimizes whole organisms to their ecological niches [38], pp. 526–527 and [46], pp. xxxv–xxxviii. Particularly, Dobzhansky [31], p. 9 conceptualized an adaptive landscape with peaks representing species, and natural selection causing peak climbing via an increase of adaptive value. Dobzhansky's adaptive landscape allowed a smooth extrapolation from microevolution to macroevolution because it suggested that complex traits originate for an ultimate purpose. Likewise, the second novel theory applied natural selection directly to long-preserved units, like species or genes [38, 39]. This allowed a link between microevolution and macroevolution because it suggested that complex traits are selected for their ultimate effects on long-preserved units.

Microevolution thus became the whole truth of evolution for two reasons. First, Dobzhansky [31] included in "genetic principles" natural selection but not other deterministic forces. This led to a focus on the genetics of microevolution [36, 37, 44, 47, 48], and distracted away from exclusively macroevolutionary forces and processes. Second, evolutionists extrapolated from microevolution to macroevolution, with the idea that complex traits originate for the ultimate effect of maximizing fitness or ensuring the survival of long-preserved units [31, 38, 39]. In this latter endeavor, however, evolutionists replaced a Divine Creator with Natural Selection as a teleological force [49], p. 2; [50–52]. In making natural selection a teleological force that operates for final causes [49], however, evolutionists violated the core tenets of Darwin's theory that natural selection works only for immediate benefits and through gradual stepwise processes [9, 12, 15].

## *Theory*
### Natural reward as force

Despite the inherent contradiction within their theory, evolutionists expressed extreme confidence in their theory. Stebbins [44], pp. 305–306, writing on behalf of prominent evolutionists, stated they could not imagine the appearance of new facts that would overthrow their conclusion that macroevolution is simply microevolution over long time periods (among other conclusions). However, Darwin's theory assumed resource limitation, and the vast majority of the sun's power, which historically fueled the expansion of life on earth, remains unexploited by life (the sun emits $3.7 \times 10^{26}$ W [53] and only $1.8 \times 10^{16}$ W strikes the earth's outer atmosphere [54]). The sun is, moreover, by modern estimates, just one of a trillion stars in the Milky Way [55], and the Milky Way is but one of two trillion galaxies in the observable universe [56]. An exhaustive discussion of how much energy life on earth uses relative to that which strikes the earth is unnecessary to realize that the total state of nature is a vast overflow of free energy rather than resource limitation. Thus, Darwin's framework (Fig. 1) is not the only one possible. An alternative framework has an assumption of resource abundance at its base (Fig. 2).

The skeletal outline of an alternative theory is simple (Fig. 2). Resource abundance rewards those that are the first to invent or disseminate traits required to exploit those resources. Thus, resource abundance leads to a struggle for economic power or dominance sensu [57], p. 207; [58], p. 199. I call this struggle *the struggle for supremacy*; see also [59], p. 226. Under the struggle for supremacy, the first forms to innovate are naturally rewarded with monopoly profit. The inevitable consequence of natural reward is an expansion of life to utilize new resources, and thus growth of the natural economy. Over vast time scales, as life expands or previous life forms go extinct, natural reward favors forms more capable of originating and disseminating novelties allowing resource capture. I argue that *natural reward* is a force of nature separate from natural selection, and that natural reward leads to the increased innovativeness or *advancement* of life with time.

What form of competition ensues when organisms are competing to exploit untapped resources? To answer this question, I note that the first form to exploit a wholly new resource zone will have a first-mover advantage, which prevents other forms from evolving into those places [32, 33, 60–65]. For example, the first form that expands to utilize a new resource will typically gain an incumbent advantage [61, 66]. This prevents other forms from radiating into the resource base [61], p. 206; [67]. Therefore, competition ensues as a race to *innovate*, where *innovate* means *invent* and *disseminate* [68]. Thus the struggle for supremacy differs from the struggle for existence because it is a race to be first, in which the winner takes all (Table 1). It is also a competition to find new resources, rather than better use those already in use.

What might we call the form of competition that occurs under the struggle for supremacy? Most authors refer to macroevolution as involving *passive replacement* rather than *active replacement* [66, 69, 70]. Benton [66], p. 188 used the term *weak competition* to refer passive replacement. In contrast, I use the term *indirect competition* for two reasons. First, there is nothing weak about the struggle for supremacy. The struggle for supremacy is a struggle for economic power. Second, Darwin [1] typically invoked a form of direct competition in macroevolution, in which the species or higher taxa in competition are simultaneously in the same location and competitively exclude each other. Thus, the term *indirect competition* contrasts to *direct competition* because it does not assume the higher groups in competition simultaneously reign supreme; rather, they are in a race to conquer (this use of *indirect competition* is not to be confused with *apparent competition* in ecology, nor is resource abundance on a large scale to be confused with a temporary overflow of nutrients or food).

To pay tribute to the importance of the struggle for supremacy, I identify natural reward as a second deterministic force of evolution. I use the term *reward* because it is often used to denote the supernormal monopoly profits earned by innovators in human economies (Table 2). To identify natural reward as a force, I dispense with the convention, implied by the neoDarwinian theory, that the environmental and physical causes of differential fitness are the forces of evolution that cause natural selection [71], pp. 13–17, 31. Instead, I assume natural selection and natural reward are forces of nature that both manifest their effects through differential organismal reproductive success. I refer to natural reward as a *deterministic* force because it, like natural selection, is non-random and assumes that the causes of events are antecedent. For example, natural reward assumes the causes for origin of inventions come before the causes for success.

In relaxing the assumption of resource limitation, it is important to keep in mind that in physical terms, the earth is an open thermodynamic system, which draws energy from outside [72]. Although most proximate opportunities for expansion will stem from the opening of new niches or extinctions on earth, ultimately the untapped energy from the sun allows opportunities to continually arise. Moreover, the only thing holding life back from nearly indefinite expansion is its ability to innovate. For example, the vast majority of the sun's power remains unexploited because no life form has yet invented or disseminated the novelties required to more fully exploit it [73]. At the same time, this also means that the power of population alone is insufficient to ensure resources are exploited. Organisms must invent or disseminate the novelties necessary to exploit novel resources, otherwise those resources will remain unexploited.

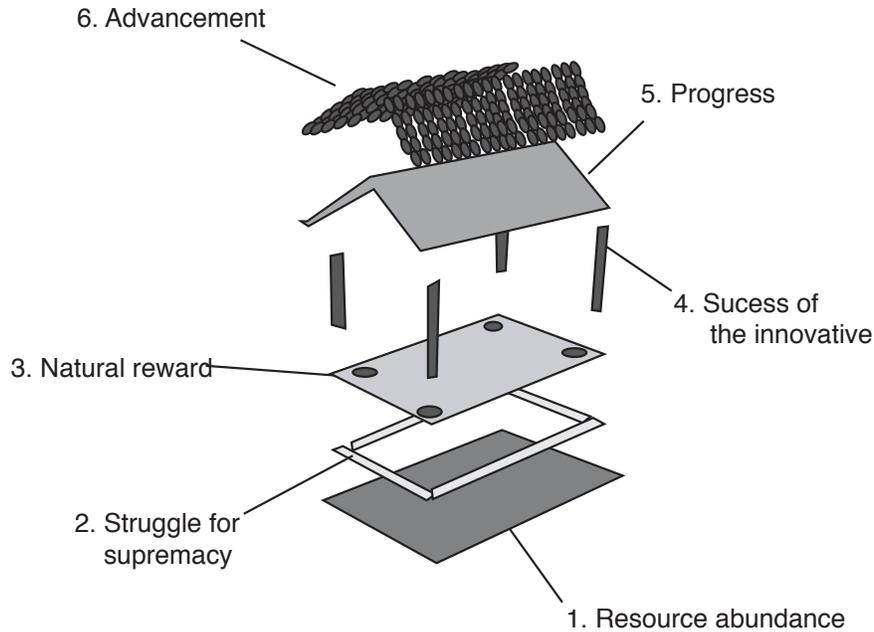

**Figure 2. Framework of the Theory of Natural Reward.** The theory of natural reward assumes a similar logical structure of Darwin's framework, except with an initial assumption of 1. resource abundance rather than resource limitation. Everything else follows.

I therefore conceptualize the struggle for existence as occurring between alleles within species. This is consistent with the standard neoDarwinian conceptualization of microevolution as allele frequency change [74–76], and Darwin's argument that the struggle for existence is most severe within species [1], pp. 68, 75, 76, 121, 320, 468. I conceptualize the struggle for supremacy, in contrast, as a race between gene complexes. The gene complex that wins the race to innovate will give rise to a higher taxon and may be revealed in its core genetic regulatory network [77]. The loser will not be seen. Thus, competition is between potential alternatives, and these alternatives represent gene complexes (Table 1). Natural selection thus results in survival of the optimized alleles [26], while natural reward results in success of the innovative gene complexes (Table 1). The question posed to the theory of natural reward is how these forces interact to determine the grand patterns of evolution.

## Theory of natural reward

Just as Darwin was led to appreciate the importance of natural selection by analogy to the economic theories of Malthus, we may appreciate the importance of natural reward by analogy to modern theories of macroeconomic growth [78–81]. By analogy, natural selection plays a role similar to an inventor in human economies, while natural reward plays the role of entrepreneur. Particularly, natural selection gradually tinkers [82] contraptions through a freeflow of creativity. Natural selection often combines old things in new ways, and gives familiar materials new and unexpected functions. Natural selection has no regard for long-term growth or market share and uses whatever materials are available to create workable objects [82]. Natural reward, in contrast, places useful inventions originated by creative tinkering in unexploited markets before potential competitors. By analogy to humanity's most profitable company, natural selection plays Nature's "Woz," and natural reward plays Nature's

**Table 1. The struggles of life.**

|  | Struggle for Existence | Struggle for Supremacy |
|---|---|---|
| Resources | Limited | Unlimited |
| Resource use | Use resources already in use | Expansion to new resources |
| Typical form of competition | Direct, between co-existing alternatives | Indirect, race to be first |
| Typical levels of competition | Alleles within species | Gene complexes shared by species and higher taxa |
| Typical outcome of competition | Survival of the optimized | Success of the innovative |

*The term *optimized* is preferred over *fittest* (Fig. 1) for two reasons. First, it does not result in confusion between *adaptedness* and *reproductive success*. Second, *optimized* makes a sharp distinction to *innovative*.

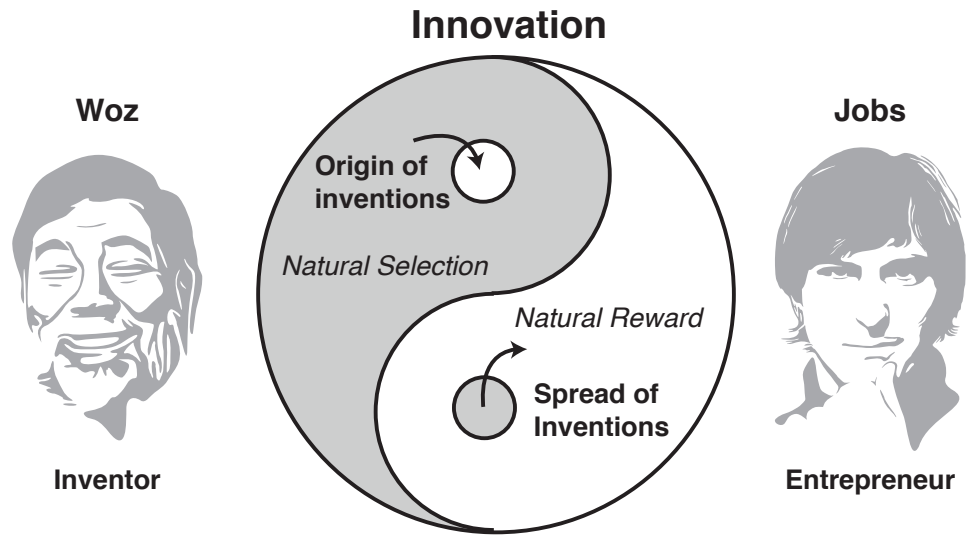

**Figure 3. The components of innovation.** Under the theory of natural reward, by analogy to humanity's most profitable company, natural selection plays the role of Nature's inventor (Woz), while natural reward plays the role of Nature's entrepreneur (Jobs).

"Jobs" (Fig. 3). Whereas Woz invented contraptions for non-entrepreneurial reasons (e.g., to impress his friends [83]), Jobs envisioned emerging markets and directed the fine tuning of Woz's inventions to achieve great economic power [84, 85]. A similar distinction between the inventor and entrepreneur is illustrated by the roles of the McDonalds brothers as inventors and Ray Kroc as the entrepreneur in the origin and spread of the McDonald's franchise [86].

Because the theory of natural reward depends on a distinction between the inventor and entrepreneur, I provide a simple definition of *entrepreneurship*. The term *entrepreneurship* has long escaped a simple definition because of all the various tasks that human entrepreneurs must fulfill. Human entrepreneurs must start companies, organize means of production, break down resistance, advertise their products, undertake ventures, take risks, and expose themselves to uncertainties of prices [93, 94]. I define *entrepreneurship* as the act of disseminating inventions. This definition of *entrepreneurship* has three attractive qualities. First, it allows a link to nature where the entrepreneurial role of spreading inventions also occurs. Second, this definition of *entrepreneurship* allows for a simple definition of *innovation*. If entrepreneurship is the act of disseminating inventions, then *innovation* is the combination of invention (the act of inventing) and entrepreneurship (Fig. 3). Everything else carried out by human entrepreneurs is incidental. Third, this definition of *entrepreneurship* allows for a simple prediction. In human societies, the successful entrepreneur not only makes himself rich; he also makes the inventors with whom he is associated rich (assuming enforcement of patent laws). Steve Jobs, by disseminating Woz's inventions, made himself and Woz rich [83]. In turn, we expect natural reward, acting as nature's entrepreneur, to favor the success of the *innovative*—and this includes both the *inventive* and *entrepreneurial* qualities in organic life (e.g., traits involved with evolvability and disseminating inventions; see below).

In making this analogy between natural reward and human entrepreneurs, I do not imply that gene networks have foresight, as human entrepreneurs sometimes do. Instead, I suggest that gene networks evolve to behave *as if* they have foresight. Two forms of foresight may, moreover, be expected. A first form is that which allows organisms to anticipate specific threats experienced by the species in the past. An example of this form of apparent foresight is CRISPR-Cas mechanisms of immunity in bacteria [95]. This allows an organism to anticipate a threat that its ancestors previously encountered [96]. Such a foresight behavior could originate by standard microevolution by natural selection for that purpose, because it involves adaptation to a particular niche. It may also be favored by natural reward, if it allows population expansion. A second form of apparent foresight allows individuals in a species to exploit new niches, not previously encountered by the species. This form of foresight may not be expected to originate by natural selection for the purpose of exploiting new niches; however, it may be favored by natural reward for this reason. This second form of apparent foresight reflects the intangible quality of successful human entrepreneurs. Because new markets are difficult to predict, this latter form of apparent foresight may involve general strategies for adapting, exploring, and discovering, rather than specific anticipated responses ([90, 91] and Table 2).

If natural selection and natural reward can both operate through the intermediary of organismal reproductive success, how do they interact to produce the grand patterns of life's history? To answer this question, I first establish assumptions on the units of evolution. Under the theory of natural reward, the unit of force is the gene complex, the unit of reproduction is the organism, and the unit of competition is the allele or gene complex. Thus, a force, natural selection or natural reward, operates on a unit, normally the gene complex, and when the power of the force overpowers random effects determining organismal reproductive success, the optimized alleles survive or innovative gene complexes succeed.

How do natural selection and natural reward interact? I suggest they interact in two ways. First, through a process of invention-conquest macroevolution, which may occur on a time frame of 1 to 10 million years for most macroorganisms [63], they fill new ecological niches. Because the only way to avoid extinction in the long term is to increase population size, I refer to such population

**Table 2. Examples of terms relevant to the theory of natural reward (bold).**

| Person | Statement | Comment | Refs |
|---|---|---|---|
| Queen Elizabeth I | At his suit: it is right that **inventors** should be **rewarded** and protected against others making profit out of their discoveries. | Early law conveying a reward for innovation in human culture *. | [87] |
| R. Cantillon | Then the Merchants or **Entrepreneurs** of the Market Towns will buy at a low price the **products** of the Villages and will have them carried to the Capital to be sold there at a higher price: and this difference of price will necessarily pay for the upkeep of the Horses and Menservants and the profit of the **Entrepreneur**... | First use of the term *entrepreneur* also highlights the basic function: to spread inventions. | [88], p. 151 |
| J. B. Say | In manufacture…if success ensue, the **entrepreneur is rewarded** by a longer period of **exclusive advantage**, because his process is less open to observation. In some places, too, the exclusive advantage is protected by patents of **invention**. | Establishes a link between the entrepreneur and the reward for innovation. | [89], p. 84 |
| J. Schumpeter | We have seen that the function of **entrepreneurs** is to reform or revolutionize the pattern of production by **exploiting an invention** or, more generally, an untried technological possibility for producing a new commodity or producing an old one in a new way … This function **does not** essentially consist in either **inventing** anything or otherwise creating… | Distinguishes the role of the inventor and entrepreneur, and explains what the entrepreneur does. | [80], p. 132 |
| | …there is or may be an element of genuine **monopoly** gain in those **entrepreneurial** profits which are the **prizes**. | Suggests that the entrepreneur's goal is to gain monopoly profits. | [80], p. 102 |
| F. Knight | …the **innovator** himself cannot predict the results in advance, or even be sure that the innovation will not be a failure, and consequently the activity is connected with "risk-taking"… the **entrepreneur** is simply a **specialist in risk-taking** or uncertainty bearing. | Recognizes that entrepreneurship usually involves risk. | [90], p. 129 |
| C. Christensen | Markets that do not exist cannot be analyzed: Suppliers and customers must discover them together. Not only are the market applications for disruptive technologies unknown at the time of their development, they are unknowable. The strategies and plans that managers formulate for confronting **disruptive technological change**, therefore, should be **plans for learning and discovery** rather than **plans for execution**. This is an important point to understand, because managers who believe they know a market's future will plan and invest very differently from those who recognize the **uncertainties of a developing market**. | Explains general strategies of entrepreneurs that may help manage risk effectively. | [91], p. 129 |
| Aghion and Howitt | **Innovation** is a vital source of **long-run growth**, and the **reward for innovation is monopoly profit**, which comes from being able to do something that your rivals haven't yet been able to match. | Statement capturing the link between the reward for innovation and economic growth. | [81], p. 7 |
| M. Sanders | If commercialization is not assumed to be trivial and the **entrepreneur** captures the rents of **commercialization**, these rents are a **reward** for bearing the risk and making the effort to turn an opportunity into a product and organizing production. | Highlights the link between entrepreneurial risk and reward. | [92], p. 341 |

*Laws are necessary in human culture because imitation or theft of invention is possible [89], p. 131.

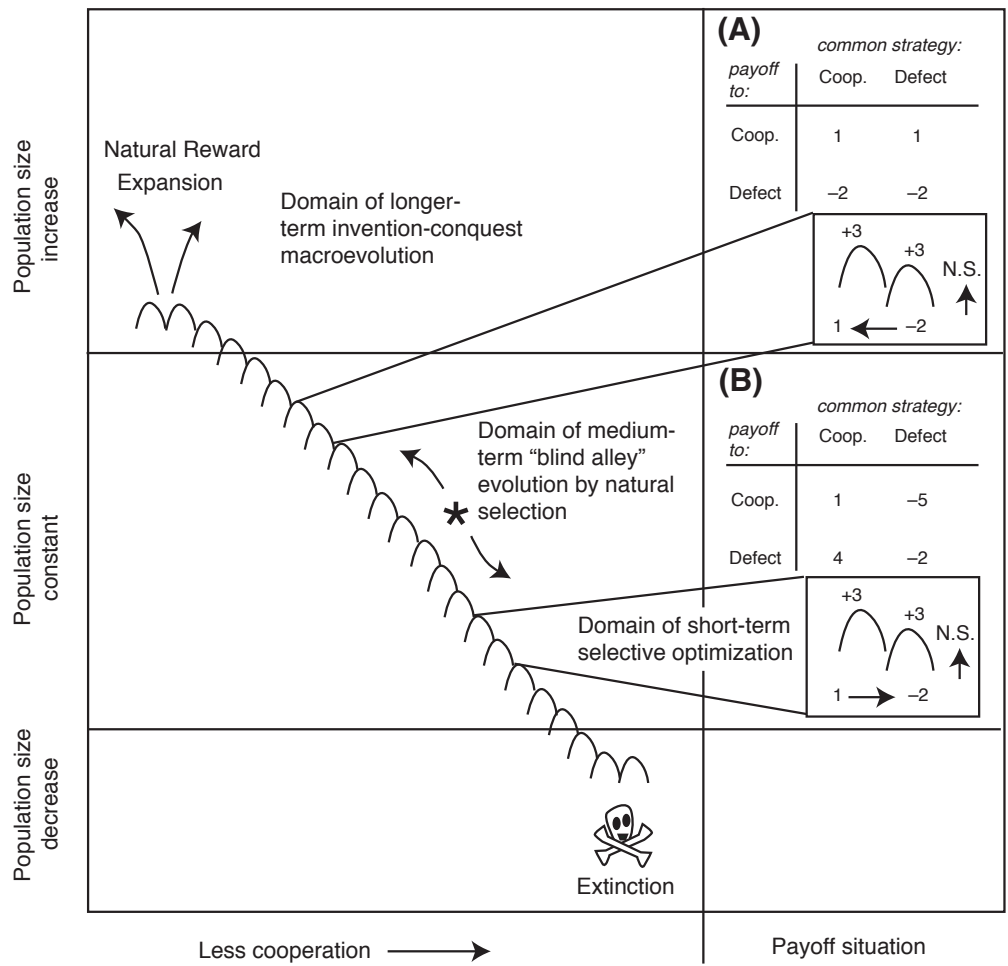

**Figure 4. Invention-conquest macroevolution.** Natural selection can lead to extinction or expansion but natural reward only favors expansion. A two-dimensional innovation landscape in which the Y axis represents three discrete states of population growth and the X axis represents the amount of cooperation. At point marked by an asterisk, the population may increase in cooperation by moving to the left, or decrease in cooperation by moving to the right. Moving to the left will lead to the origin of novelties and population expansion. Moving to the right will lead to the loss of novelties, population decline and extinction. Which occurs depends on the population structure. **(A)** The population is clone structured, meaning that individuals only encounter clonemates during the part of the life cycle when cooperation is possible (but mating is random). Payoffs to alleles coding for indiscriminate strategies on the left (rows) in populations where strategies on the right (columns) are common. Here, alleles coding for cooperation receive a cooperate-cooperate payoff, and alleles coding for defection receive a defect-defect payoff. The evolutionarily stable strategy (ESS) is then to cooperate despite the prisoner's dilemma at the level of individuals (note: payoffs are to alleles, and depend on interactions between individuals. Interactions between individuals of different strategies do not occur in clone-structured populations, so allele payoffs can be different than individual payoffs. The payoff for alleles in **B**, which conforms to a standard prisoner's dilemma, applies to individuals in **A** and **B**). **(B)** The population is well mixed so that clones frequently interact with other clones. Here, alleles coding for defection will sometimes gain a defect-cooperate payoff, and alleles coding for cooperation will sometimes receive a cooperate-defect payoff. Consequently, the ESS is defect. With every evolutionary step, natural selection (N.S.) increases gene optimization value +3, independently of whether the population is heading toward expansion or extinction (see payoffs at right).

expansion as *progress*. I use the term *progress* because long-term population expansion decreases the probability of extinction relative to population decline [1, 97, 98]. At the very minimum, the basic point of life is to avoid extinction and so population expansion is progressive. For those who argue the basic point of life is to expand [57, 58], the same definition of progress is relevant. Exceptions may be meaningful only for entities that have transitioned to immortality or invincibility, or for those who believe the purpose of life is to go extinct. Because there is no evidence of the former and the latter should probably not exist (though it might), I proceed with a notion of progress as expansion. This also conforms to common opinions on what is meant by "economic progress", though short-term growth of human economies should not be viewed as progressive if such growth leads to long-term decline or extinction.

Second, through a process of extinction-replacement megaevolution, which occurs on the order of 50 to 150 million years for macroorganisms [99, 100], natural reward and natural selection interact to release constraints on progress and increase the overall innova-tiveness of life. By analogy, a scientific system *progresses* when it invents models or makes key discoveries that increase the total domain of knowledge, e.g., every 5–20 years; but it *advances* when a new core theoretical framework allows the theory to become a more potent engine of exploration and discovery, e.g., every 50–200 years [92, 101]. Before getting to this prediction for advancement, however, I will first focus on how invention-conquest macroevolution results in progress.

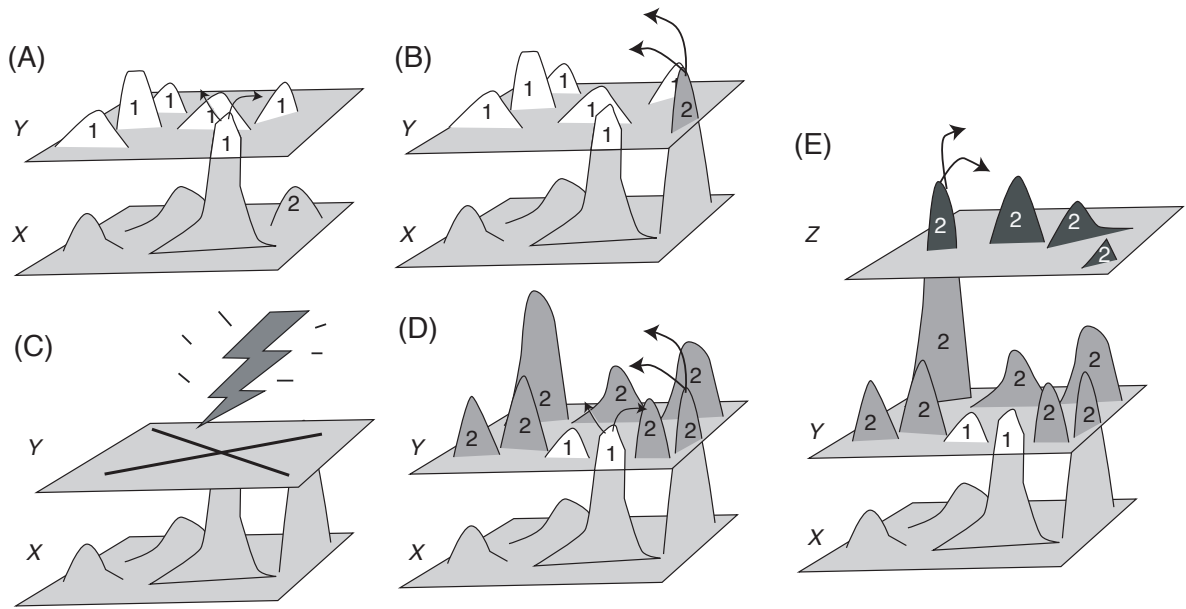

**Figure 5. Extinction-replacement megaevolution.** A three-dimensional innovation landscape in which **(A)** in the initial state, lineage 1 wins the race to innovate over lineage 2 and conquers a new biological market, going from tier X to tier Y. **(B)** Lineage 2 now achieves the ability to expand into tier Y, but cannot because of the monopoly of lineage 1 already present. **(C)** Environmental change and mass extinction kills all organisms in tier Y, but lineage 2 remains poised to reinvade tier Y. **(D)** Competing head-to-head, lineage 2 expands faster than lineage 1 and conquers tier Y, also producing more species. **(E)** Lineage 2's invasion of tier Z reveals another tier Z that was never exploited by lineage 1.

To explain how natural reward produces progress, I first give a two-dimensional innovation landscape in Fig. 4. This figure focuses on cooperation as a model. I focus particularly on cooperation because at each level of biological organization—unicellular, multicellular, and colonial—cooperation among close relatives helped facilitate the origin of other innovations, including those involved with transmitting information across generations [102–105], and various other traits allowing expansion into new habitats [106–108]. The cooperation among cells was necessary for the evolution of specialized cell types, the evolution of diversified organs, and multicellular organisms of large size and complexity [109]. A high degree of cooperation and mutual cohesion at one level was, moreover, a necessary prerequisite for inclusion into yet higher levels of hierarchical organization [110]. Non-cooperation, in contrast, does not have such long-term effects on expansion. Thus, cooperation is a useful model because it was consistently a major innovation, leading to some of the biggest ecological expansions in the history of life on earth [111, 112], while its alternative, non-cooperation, was not.

In a case where a population is kin structured, natural selection can easily favor cooperation as shown by the payoff matrix in Fig. 4A. This payoff matrix is derived from a standard prisoner's dilemma, where high relatedness among associates restricts help to those sharing the same genes. In a well-mixed population, where individuals often associate with nonkin, however, natural selection easily favors defection in the absence of other mechanisms maintaining cooperation (Fig. 4B). In cases of high mixture, cooperators interact with non-cooperators, in which case non-cooperators are favored for avoiding the costs of helping others. Thus, natural selection can easily favor either cooperation or defection. However, natural reward only favors cooperation as a key innovation (Fig. 4). Thus, only natural reward exerts an asymmetric effect on expansion and progress via cooperation.

Fig. 4 thus has three important points. First, it suggests that natural selection optimizes only with respect to the extreme short term, as represented by minor peaks on an overall slope. This keeps in line with Darwin's argument that natural selection works only through slight and successive modifications. It also distinguishes Fig. 4 from previous adaptive landscapes, because it suggests that the larger peaks do not represent selective optimization. Second, Fig. 4 shows that natural selection is a blind process that has no regard for future benefits of populations. Natural selection can doom populations to extinction in both natural [113, 114] and laboratory environments [115, 116], for example by overspecializing species [98, 117, 118], or by favoring cheating (Fig. 4 and [116]). Third, although evolution by natural selection will not always just as easily lead to extinction as success, as depicted in Fig. 4, natural selection still is relatively blind to the long-term effects of traits it produces. Natural reward, in contrast, favors only those traits that allow expansion. In the long term, natural reward, by differentially favoring novelties that expand popula-tions [68], has a highly asymmetric effect on progress and expansion.

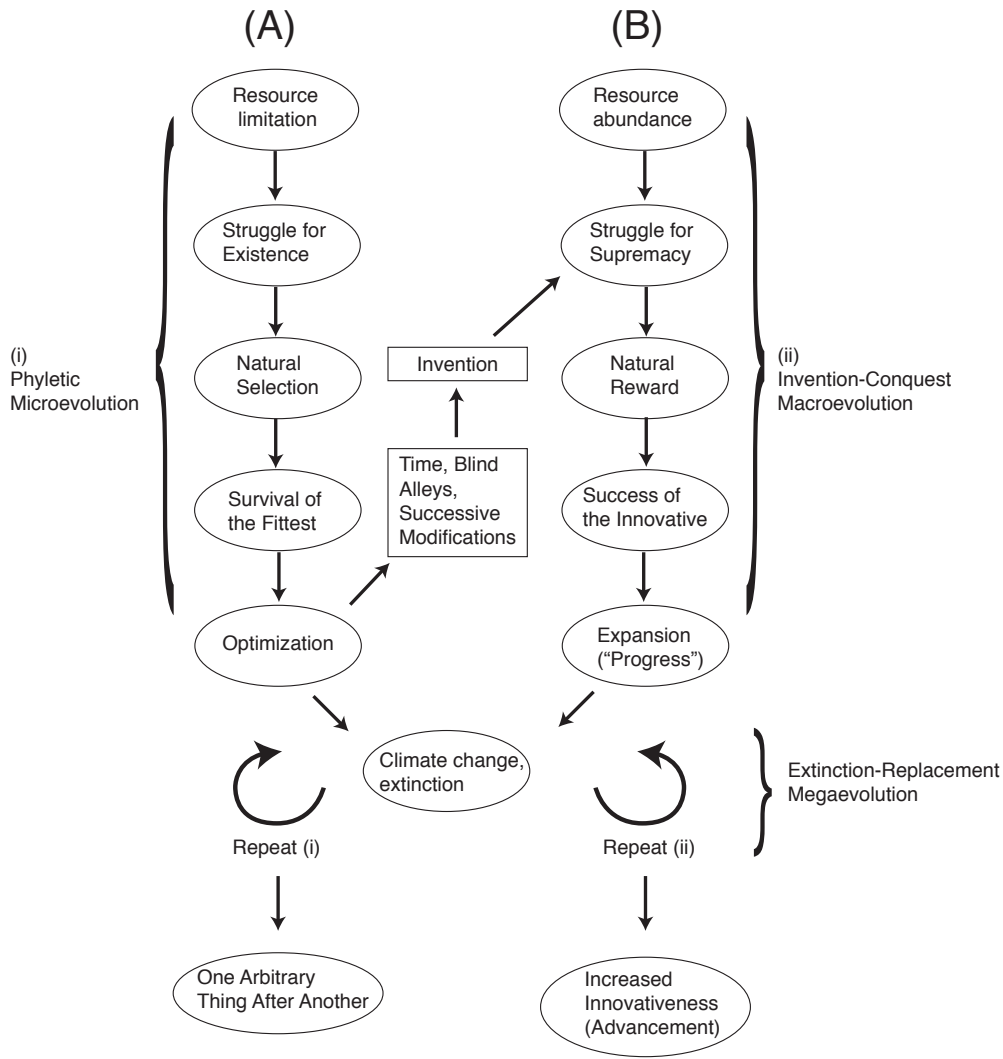

**Figure 6. Comparison of theories of natural selection and natural reward.** (**A**) The theory of natural selection includes A only. (**B**) The theory of natural reward includes A and B, as well as the connection between them.

Natural reward is a savvy entrepreneur who disseminates successful inventions to new markets. After natural reward disseminates inventions, natural selection will often fine tune inventions for their new or expanded roles. Because natural selection plays a role as originator and finisher, it might appear that all fit between organism and environment is caused by natural selection alone. It is important to remember, however, that natural reward plays a crucial role in creating a match between organism and its environment. Natural reward creates this fitness by disseminating particular traits to the environments for which they happen to be fitted to exploit. Natural selection is expected initially only to manage with available odds and ends, with no special project [82], to produce *contraptions* [119]. *Contraptions* are distinguished from *contrivances* because their various subparts were not all uniquely and expressly chosen with a particular final goal in mind, but are the result of jury-rigging of available materials [38] p. 120. A major effect of natural reward disseminating inventions is to make *contraptions* appear as if they are *contrivances*. This is important to note, because many evolutionists have taken apparent contrivance as evidence of natural selection as a teleological agent [49, 50, 120], rather than natural selection and natural reward working together (and whereas in human culture entrepreneurs often usurp credit for the inventor's work [85, 86], evolutionists often give the inventor credit for the entrepreneur's work in organic evolution instead).

The two-dimensional innovation landscape presented in Fig. 4 and the foregoing discussion explains the roles of natural selection and natural reward in facilitating expansion. However, under the struggle for supremacy, the rewards go to those who are *first* to innovate. Thus, if one lineage expands and conquers an environment, it can be difficult or impossible for another to later invade because of the incumbent advantage. If the winner is based on a fundamentally constrained genetic system, monopoly may stifle progress. How, then, are constraints on progress lifted, so that life may become more innovative with time?

I now present a three-dimensional innovation landscape. In contrast to Fig. 4, which explains only the relatively short-term process of invention-conquest macroevolution, Fig. 5 shows a longer-term process of extinction-replacement megaevolution. Fig. 5 focuses only on inventions leading uphill toward expansion, thus taking an asymmetric focus on innovation. Fig. 5, in turn, shows four things. First, the initial lineage to conquer an environment can usurp the market share and prevent invasion by others. As shown in Fig. 5A, lineage 1 is the first to win the race to innovate, while lineage 2 only later becomes capable of invading the environment denoted tier *Y*. However, the prior presence of lineage 1 prevents conquest by lineage 2 even after lineage 2 invents the novelties that would allow it to invade a vacant tier *Y* (Fig. 5B). Second, mass extinction puts potential competitors in the struggle for supremacy back on equal footing (Fig. 5C). This allows lineage 2 to conquer the initial tier *Y* after mass extinction of lineage 1. Third, some of lineage 1 carries over to fill a small region of tier *Y* (Fig. 5D). Thus, a former incumbent can retain some market share. Fourth, lineage 2 more fully fills tier *X* than lineage 1 formerly did, as represented by more peaks. Lineage 2 is also able to conquer a wholly new resource zone represented by tier *Z* (Fig. 5E).

Why would new species, reinvading following the extinction of incumbents, however, be more capable of expanding into a new tier *Z* as depicted in Fig. 5? The main reason is that any form better at dispersing and radiating into the initial tier *Y* might also be more likely to invade a novel tier *Z*. However, this raises the question of why there might be a form better able to radiate back into tier *Y*, after lineage 1 has held the monopoly for so long. One possibility is that for the period within which a particular dynasty reigns supreme, there will always be evolution going on at the marginal fringes that may be capable of producing a more innovative genetic system. After an extinction event, the existing variation in innovative capacity is acted upon by natural reward. Thus it may be generally assumed that after sufficient time, the form with the incumbent advantage may not strictly be the best—it retained monopoly only because it was first. Extinction, by leveling the playing field, reveals latent variation in innovative capacity in the whole population.

We are now in a position to compare the basic differences between the theory of natural selection and natural reward (Fig. 6). Under the theory of natural selection, which assumes a single deterministic force, the broad-scale course of evolution proceeds as repeated bouts of phyletic microevolution (Fig. 6). The necessary implication is that, even if evolution is a history of key innovations [105], the broad-scale course of evolution is one arbitrary thing after another [121]. The theory of natural reward, in contrast, provides a connection between microevolution and macroevolution by viewing natural selection as a creative process that produces complex inventions that are the fodder for the struggle for supremacy (Fig. 6). The struggle for supremacy leads to invention-conquest macroevolution, and following climate change and extinction, extinction-replacement megaevolution. Natural reward, by differentially favoring inventions that exploit untapped resources, provides directionality to evolutionary history. Over vast time frames, natural reward leads to an increase of overall innovativeness, and thus the advancement of life.

## Unifying power

A main prediction of the theory of natural reward is that traits that allow capture of untapped resources will allow population expansion. The idea that novelties allowing exploitation of new resources allow population expansion has figured prominently in former discussion of innovation, adaptive radiation, and the energetics of evolution [32, 33, 60–66, 68, 122]. The basic concept has also been demonstrated in laboratory experiments [123, 124]. Evolutionists in these cases typically assumed that novel opportunities provided a "selective pressure" for expansion. However, the typical form of competition is indirect between higher taxa, rather than direct within species [32, 33, 60–66]. According to the theory of natural reward, the force involved is natural reward. But this raises the question: why not just call it *natural selection*? What additional insights are gleaned by realizing natural reward is a separate macroevolutionary force?

A first example showing the unifying power of the theory of natural reward is why sexual reproduction is widespread [125–128]. Because natural selection works only for short-term benefits, the problem has involved identifying a short-term advantage to sex or the subcomponent traits of which sex is formed. This allows a division of the problem of sex into two parts: why sex originated [129, 130], and why sex is maintained [131–133]. The former problem is one of deep time and requires explaining the origin of cell fusion, separate sexes, diploidy, two-step meiosis, recombination, and so on [129, 130]. The latter problem requires identifying short-term selective advantage to sex that explains why sexual populations are not overtaken by asexuals. Hypotheses to explain the maintenance of sex include those focused on genetic benefits of speeding adaptation or breaking apart negative genetic associations [125, 126], specialization or division of labor [134, 135], parasite resistance [132], or purging of deleterious mutations [133]. Arguments focused on the origin or maintenance of sex, however, give the impression that some essential feature of the situation is being overlooked [136].

What is being overlooked? The theory of natural reward suggests it is the match between question and answer. The question usually asked is why sexual reproduction is widespread, and the answer given is why sex originated or is maintained [125–128]. The theory of natural reward uses natural selection to explain origin and maintenance, and natural reward to explain success. Questions of success are different, moreover, because they focus on incidental consequences. For example, sex could be widespread because of its effects on hybridization and introgression, which contribute to invasion of novel habitats [137–141]; because it allowed the spread of selfish genetic elements, which ultimately contribute to gene regulatory network evolution and novelty [142–147]; or because sex allows sexual selection, which lead to the evolution of weaponry [148–150], and (possibly) large brains [151, 152]. Any such long-term consequences of sex could lead to a greater net diversification rate for sexuals, which would cause sex to become widespread [153, 154]. Crucially, when one separates cause and effect, it is possible to entertain many possible hypotheses that could contribute to the ubiquity of sex.

Similarly, those who focused on the ubiquity of cooperation typically focused narrowly on particular genetic hypotheses for why cooperation originated or is maintained [111, 112, 155–157]. To explain why cooperation is widespread, it is important to focus on long-term effects. Cooperation may have numerous long-term effects that vary between taxa. The most important effect of cooperation may have been in allowing the evolution of heredity [103–105]. More commonly, cooperation may have allowed new interactions with the external environment that vary between taxa [106, 107, 107–110]. In social amoebae, cooperation between differentiated cells allows a stalk useful for dispersing spores [158, 159]. In vampire bats, cooperation between non-differentiated individuals allows survival through cold nights [160]. How important were these traits for ecological success? No model for the maintenance of cooperation [157] will fully answer this question. To understand why cooperation succeeds in a particular case, it is important to understand the natural history of the traits that cooperation allows, and how these traits interact with the environment [106, 107, 158, 159].

Traits enhancing the spread of inventions, likewise, were not previously included within the same explanatory framework. This is shown, for example, by a recent attempt to bring traits enhancing dispersal into the overall conceptual umbrella of modern evolutionary theory. Particularly, Shine et al. [161], argued that traits enhancing dispersal are significant because they cause changes of phenotypes in space even in the absence of differential reproductive success. They referred to this phenomenon as *spatial sorting*. However, the theory of natural reward suggests that traits enhancing dispersal are important because they disseminate inventions to exploit untapped resources, thus causing large population expansions. In this sense, their ubiquity is an indicator of natural selection's bold and entrepreneurial associate, natural reward (Fig. 3; not natural selection's "shy younger sibling" [158]). In contrast to spatial sorting, which can occur in the absence of differential reproductive success (or conceivably without reproduction at all), natural reward depends on differential reproductive success (and reproduction) of dispersing forms.

To apply the theory of natural reward to the phenomena that Shine et al. [161] describe, consider their description of cane toads with longer legs. These individuals were capable of dispersing more rapidly, but the longer legs caused spinal injuries. From the perspective of the theory of natural reward, this would suggest that longer legs would be favored under the struggle for supremacy, as long as opportunities for expansion persist, but not by natural selection under the struggle for existence, once expansion ceases. Shine et al.'s [161] review of similar traits can also be seen as collecting traits that are differentially favored by natural reward (e.g., behaviors causing individuals to attach themselves to other species like burrs and ticks, seeds and fruits in digestive tracts or human vehicles, bold risk-taking behavior, and the ability to handle the stresses of intense physical activity [161]). Shine et al. [161] did not see the more general importance of dispersal traits, nor did they unify them with evolvability traits, because they did not realize that natural reward could operate through differential reproductive success to favor expansion.

More generally, under the theory of natural reward, the role of entrepreneurship, or disseminating inventions, is not restricted to traits allowing dispersal through space. Natural reward can also favor traits allowing dispersal through time. Examples of time-dispersal traits include hardy cysts, seeds, or diapause stages, or stochastic phenotype switching traits [162–166]. Models show that time-dispersal traits can allow survival through major environmental catastrophes and thus differential population expansion [165–168]. Another trait that may allow dispersal through time is facultative parthenogenesis based on cues like the absence of mates or stress [169–172]. This ability to reproduce asexually based on cues could allow species to survive major catastrophes, like floods or volcanic eruptions [173].

Natural reward may also favor traits that allow spatial and temporal diffusion of novelties, but which are not classic space or time dispersal traits. One such trait is the near universal genetic code [174]. One scenario to explain the initial fixation of a single genetic code is that some life forms specializing on the use of an initially abundant code gained access to larger pools of protein innovations [175]. The competition between various life forms initially specialized on different codes may thus have been a race to innovate, which rewarded those that were the first to have specialized on a code that was both abundant and relatively useful. In a similar way, a scenario to explain the near fixation of a single language for scientific communication is that scientists using a common language gained access to larger pools of discoveries and methods [176, 177]. This could be true even if the language chosen was not optimal or even the best among alternatives (e.g., English has poor tracking between words and numbers [178, 179]).

By drawing attention to the importance of the disseminating inventions on expansion, the theory of natural reward unifies standard evolvability traits, involved with the origin of inventions, with traits involved with the spread of inventions. This unification also helps show how some simulation studies could be taken to support either. Lehman and Miikkulainen [180], for example, showed that if a trait increases the probability of exploiting a new niche, it can expand as a consequence of repeated bouts of extinction-replacement megaevolution. Although they took their results to suggest an increase of evolvability, their simulation can also be taken to predict an increase of dispersal traits as well. At least one prior analysis included both sorts of traits in the same review [181], consistent with the theory of natural reward.

What are some examples of extinction-replacement megaevolution in nature, leading to advancement? One well-known example is the initial radiation of the dinosaurs to fill terrestrial environments in the Mesozoic, followed by replacement by the mammals in the Cenozoic. The initial radiation of dinosaurs would be reflected by Fig. 5A, first starting 240 million years ago [182]. By around 200 mya, however, the mammals had appeared and were beginning to evolve the traits that would contribute to their later diversification and success [183–185]. It was not until 65 mya, however, that the mammals got their chance, following the Chicxulub asteroid impact, and a series of volcanic eruptions that annihilated terrestrial animals weighing > 50 pounds [186]. At this time, mammals had attained some of the general adaptations [183–185] that helped them in a struggle for supremacy with the remaining dinosaurs. With the large dinosaurs gone, the mammals radiated into the open niches for large body size [187] as suggested by Fig. 5D. The dinosaurs did, however, retain some market share as birds diversified, analogous to the white peak on tier $Y$ in Fig. 5D. The mammals also yielded greater exploitation of marine environments, as suggested by the diversity of cetaceans [188] compared to marine reptiles at the K/T boundary [189].

Why would the mammals, reinvading following the extinction of large dinosaurs, be more innovative? One possibility is that for the period within which dinosaurs ruled the earth, mammals were evolving various traits that would fortuitously give them a leg up on dinosaurs in both surviving a catastrophe and radiating back into available niches (dentition, endothermy, intelligence, etc. [183, 185, 186, 190]). A resulting greater innovative capacity may have increased the likelihood that mammals repopulating an environment would break into a yet additional tier (Fig. 5E). Mammals indeed invaded a wholly new resource zone represented by tier $Z$, allowed by the human invention of language [105], mathematics, agriculture, science, and economic systems that promote growth [191, 192]. In agreement with the theory of natural reward, much of human economic growth is driven by technological innovation [81, 193]. The most innovative strategies employed by humans are also flexible methods that allow exploration and discovery into new fields and markets [91, 194, 195].

More generally, natural reward, acting through the process of extinction-replacement megaevolution, is expected to increase innovativeness. The strategies expected cause individuals to behave as if they know that new opportunities are virtually unpredictable, and thus that the best strategies for exploiting opportunities are geared for evolving, exploring, and dispersing, rather than strategies for anticipating particular threats. This form of apparent foresight deals with extreme uncertainty and the problem of predicting emerging markets [91], p. 148; [93], p. 7; [196]. It may be likened to a physical process that maximizes the future freedom of action [197], and general scientific methods that rely on primitive but flexible methods [194, 195]. Such apparent foresight strategies could be those that deal with either the origin or dissemination of inventions (Fig. 3). They might thus include general strategies that enhance evolvability or contribute to the dispersal or dissemination of inventions through space or time.

## Anomalies

Here I will explain how the theory of natural reward helps dissolve three previous anomalies of evolutionary theory. These include punctuated equilibrium, the increase of complexity and diversity with time, and the increase of energy flux or power. The first is the punctuated equilibrium [38, 198], or the sudden appearance of higher taxonomic categories in the fossil record. To explain this phenomenon, Simpson [32] hypothesized that higher categories arise following major genetic changes in small, isolated populations. These major genetic changes produce the novelties that allow invasion of new adaptive zones and speciation. In a review of Simpson's work, Wright accepted Simpson's argument [199], but later rejected it in favor an alternative that assumes novelties originate gradually [174–176]. In Wright's argument, punctuated equilibrium occurs without saltational origin of evolutionary innovations. Wright's gradualistic alternative agrees with the theory of natural reward.

Wright [174–176] argued that important novelties evolve gradually and spread suddenly. According to Wright, sudden spreading is caused either by the origin of a key novelty or movement into a new place allowed by migration or extinction of an incumbent. In the former case, there is an important role for a final minor change pushing a species beyond a threshold necessary to enter some new adaptive zone. In the latter case, there is an important role for effects of a complex trait that have nothing to do with the cause for origin. Simpson accepted Wright's hypothesis and thus assumed breakthroughs or "quantum shifts" were not saltational as Simpson [32], p. 350 originally argued, involving major change concentrated in speciation events. Instead, they involved the same type of normal phyletic evolution that occurs at other times. According to Simpson [33] pp. 350, 389, what is unusual about traits leading to the origin of higher categories is their long-term *consequences*, not their mode of origin, which is what Wright argued. Wright's argument is consistent with the theory of natural reward, because it does not assume that key innovations originate suddenly as Simpson's [32], Mayr's [203], and Eldredge and Gould's hypotheses did [204].

The theory of natural reward supports Wright's argument because it distinguishes the roles of natural selection and natural reward (Fig. 3). This allows complex traits to originate without saltation and yet have differential long-term impacts. Shifting rates of evolution are explained because the sudden expansion leads to increased rates of evolution as new forms adaptively diversify. This also helps explain why new forms often appear suddenly in the fossil record. The lineages evolving the novelties are at low density prior to crossing a threshold or extinction of an incumbent, thus leaving few fossils. Once a particular threshold is reached or extinction occurs, invasion of a new adaptive zone is possible. The drastic increase in total numbers that accompanies major innovation or extinction means that the newly abundant type will more likely be detected in the fossil record. The finding that novelties originate gradually and yet also lead to punctuated bursts of adaptive radiation, with a strong role for incidental effect, is also witnessed in technological evolution [198], p. 6.

How does the theory of natural reward bear on historical increases of hierarchical complexity [109] and ecological diversity [205]? Previously, some suggested that these increases were largely passive, as encapsulated by a "passive diffusion from the left wall" [206, 207], or "zero evolutionary force law" [208]. Others suggested a more active process that required overcoming constraints, as encapsulated by "a repeated scaling to the right" [209]. The theory of natural reward is more consistent with the second explanation, because it suggests that organisms had to invent novelties that allow capture of novel resources. In contrast to the "constraints from the right" perspective, however, the theory of natural reward identifies an evolutionary force that is involved with expansion. As applied to the origin of ecological diversity, the theory of natural reward suggests that opportunities continually arose throughout the history of life when organisms invented new ways to exploit resources. Each new innovation also typically had spillover effects. For example, each step in hierarchical organization opened up adjacent niches [210] for organisms of the same or lower degrees of hierarchical organization (e.g., parasites, mutualists and commensalists). The effect of extinction may have been to sometimes allow life to overcome plateaus of maximal diversity [205]. Although each small step in hierarchical complexity and ecological diversity can be explained by opportunities and changes on earth, overall the upward trend required new opportunities for novel resource capture allowed by copious energy from the sun.

Another question is why life has increased in total energy flux or power [211]. According to Lotka [122], life should increase in energy flux because forms that achieve higher energy flux will beat those with lower energy flux in direct competition. Thus, Lotka's *maximum power principle*, as sometimes expounded, may be tested by thrusting different sets of species into a direct struggle [212]. At a macroevolutionary scale throughout the history of life, however, such direct struggles are likely to have been of relatively minor importance compared to indirect struggles [32, 60]. An alternative explanation is that as life increases in size and complexity, those at higher rungs of the food chain draw upon a larger total amount of energy stored up in the biomass accumulated at lower levels. In that case, an overall increase of energy flux is simply an incidental consequence of life's upward expansion to higher levels of complexity and not a result of direct competition having dominated macroevolution. However, occasional direct competition with merging of continents [69] could reinforce the overall increase by increasing energy flux at a particular hierarchical or trophic level [212].

## Reversal of teleology

Williams [213], a staunch critic of teleology pp. 21, 128, recommended that a trait be assumed to have originated by selection for a purpose if it appeared designed for that purpose [213], pp. 8–9. Perhaps the most shocking prediction of the theory of natural reward is how it contradicts Williams' paradigm. The theory of natural reward suggests that apparent design in nature, as well as fit between organism and environment, is not exclusively caused by natural selection adapting species to those ends. Instead, it is evidence of natural selection going down blind alleys and creating contraptions without regard to future effects, and of natural reward seizing upon and spreading useful inventions. Natural selection at the end may be responsible for fine tuning and specializing adaptations, but much of the fit between organism and environment is caused by natural reward.

What evidence supports this role for natural reward? The first line of evidence comes from computer simulations. These simu-

**Figure 7. Crystallized NeoDarwinian framework.** Compared to Darwin's framework (Fig. 1), the NeoDarwinian structure combines resource limitation, the struggle for existence, and the survival of the fittest into a single initial concept of differential reproductive fitness. It thus promoted theories suggesting that complex traits originate for the ultimate purpose of reproductive fitness maximization or selective benefits to long-preserved units (italics). In that case, natural selection is a teleological force that explains the major trends of macroevolution (italics).

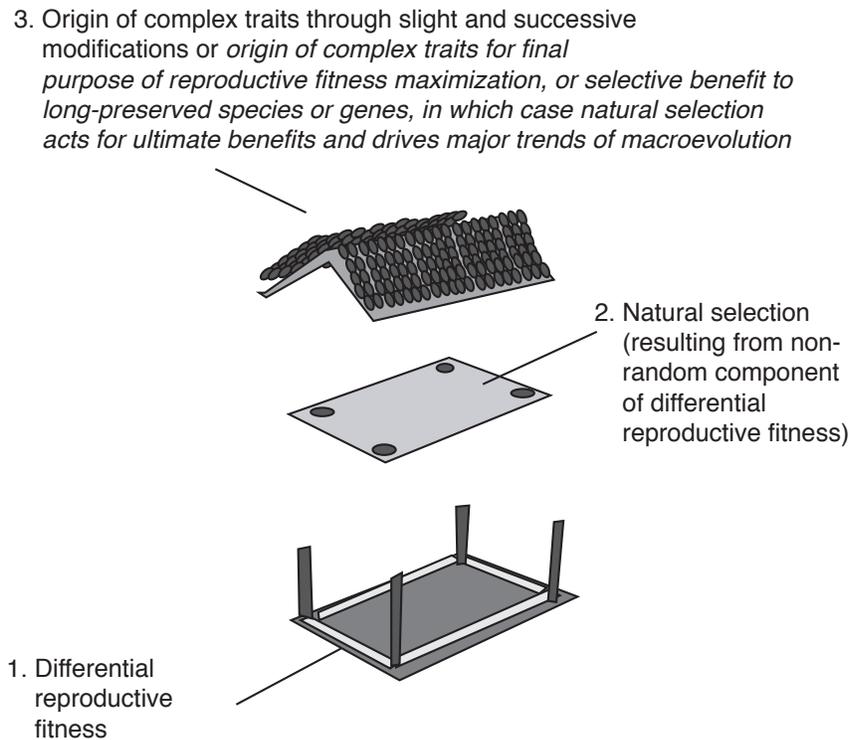

3. Origin of complex traits through slight and successive modifications or *origin of complex traits for final purpose of reproductive fitness maximization, or selective benefit to long-preserved species or genes, in which case natural selection acts for ultimate benefits and drives major trends of macroevolution*

2. Natural selection (resulting from non-random component of differential reproductive fitness)

1. Differential reproductive fitness

lations show that the more complex a trait is, and the more features of conformity appear contrived for a common goal, the less likely the trait originated for a single final cause. The reason is the necessity of *stepping stone* traits that are necessary intermediates between simple and complex traits [123, 214–217]. As an example from nature, feathers [218], bipedal locomotion, and other traits [219], were stepping stones for wings that almost certainly originated for a reason other than flight (see also [9, 20, 23, 130] for other examples). In one particularly illuminating set of computer experiments, Lehman and Stanley [216] designed two computer algorithms: one that optimizes to the final objective, and another that is a blind novelty search [215]. They found that the more complex a trait is, the less likely it could be evolved by an optimization algorithm. Only the blind novelty search could find the very complex traits because of stepping stones [215, 216]. This illustrates that natural selection works to produce complex traits as a tinkerer who goes down blind alleys [11, 82], and is consistent with the natural reward's role as nature's entrepreneur (in spreading inventions). It also shows why Williams' [213] paradigm, now widely accepted [49, 120], is wrong: the more complex a trait is, the *less* likely it originated for its ultimate apparent purpose.

Another line of evidence for the theory of natural reward comes from studies of complex traits in nature. If the theory of natural reward is correct, complex traits will often originate for reasons that have nothing to do with their ultimate apparent purpose. An example of such a trait is histocompatibility [23, 220]. This trait appears adapted to control the spread of social parasites [110, 111, 221, 222]. However, models that break a complex trait into simpler components suggest that histocompatibility evolved through a historical process, with the final selective pressure being to avoid discriminatory conflict [23, 220]. These models suggest histocompatibility only incidentally controls social parasites [220].

A final line of evidence comes from the artificial dilemmas that arise when all fit between organism and environment is assumed to be caused by natural selection. Haldane's dilemma [223] states that if an environment changes so rapidly that the species must change in multiple characters merely to survive, then the extinction rate should be much greater. Anyone who includes in *adaptation* a connotation of *origin by selection*, however, will be stuck in a perpetual dilemma. Haldane's dilemma is solved if migration can allow organisms to continue to track their environments despite environmental change; e.g., [224–227]. In these cases, the tracking of species to their environments (without expansion) allows continued "fit" even where natural selection is not powerful enough to do so. In a similar way, natural reward (with expansion) often achieves a fit with the environment by spreading traits into the environments they are fitted to exploit.

## Constraints and carry over

The largest constraints against the theory of natural reward were modern assumptions on the requirement for evolution by natural selection. Ever since Fisher [228], it became widely assumed that the requirement for natural selection is heritable variation in *reproductive fitness*, usually calculated as some composite of survival and fecundity, which may include age structure and other factors [228–230]. This paradigm crystallized the independent assumptions of Darwin's theory (Fig. 1) into a superstructure that could not be dissociated into its component parts (Fig. 7). If heritable variation in reproductive fitness is all that is necessary for natural selection (Fig. 7; [74], p. 458; [75], p. 60; [76], p. 77), then it would appear that resource limitation and a struggle for existence are unimportant [231]. Without appreciating the importance of resource limitation (Fig. 1), however, it is impossible to determine the levels and time frames of evolution by natural selection (see introduction). It is also impossible to extend the theory of evolution to include resource abundance, the struggle for supremacy, and natural reward (Fig. 2). The first step to extending Darwinism, therefore, is to return to Darwin's core framework, which has resource limitation, rather than differential reproductive fitness, at its base (Fig. 1).

Next, it is important to understand the impact of the reproductive fitness paradigm, because it will be responsible for most of the incumbent resistance to the theory of natural reward. Much theory in biology is based on reproductive fitness [156, 232, 233], and the field of philosophy of biology was founded upon it [234], p. 1; [235], p. 141. Problems that manifest under the reproductive fitness paradigm include the *units of selection* problem [233], *the meaning of fitness* problem [156, 232], and the *utility of inclusive fitness* problem [236]. Below l explain how the theory of natural reward dissolves these problems so that sterile controversies can be avoided.

The first problem is the units of selection problem. The problem originally appeared with the new requirements for evolution (Fig. 7), which obscured Darwin's explanation for the levels and time frames of evolution. Confusion over the units of evolution stemmed from implicitly conflating the units of force, reproduction and competition in a single concept of reproductive fitness [229, 232]. Under the theory of natural reward, the unit of force is the gene complex, the unit of reproduction is the organism, and the unit of competition is the allele or gene complex. Confusion also came from applying reproductive fitness concepts to long-term evolution [71, 237, 238]. Under the theory of natural reward, this confusion is removed because it is possible to separate the roles of natural selection and natural reward originating and disseminating complex traits. One does not have to formulate a single "reproductive fitness" quantity that applies to all time frames. By invoking three levels of evolutionary units (force, reproduction, and competition), and distinguishing natural reward and natural selection, the theory of natural reward resolves the units of selection problem.

The meaning of fitness problem, in turn, originated with the assumption that there is a reproductive fitness quantity that organisms maximize [239]. Some theorists assumed the key to understanding the origin of complex traits was finding this fitness quantity [120, 239, 240]. Despite years of searching for it, however, it was never found [239, 241]. Some therefore abandoned the idea that reproductive fitness maximizes [230, 239], while others abandoned the concept of reproductive fitness entirely [242]. The theory of natural reward suggests that the concept of reproductive fitness should be abandoned for three reasons. First, Darwin had no concept of reproductive fitness, but instead used *fitness* to mean *adaptedness* [1, 232]. Second, the theory of natural reward clarifies natural selection as a short-sighted force. The idea that natural selection works for the ultimate purpose of fitness maximization [49, 120, 213, 240] came from confusion between Fisher's analogy to statistical physics with an analogy to Newtonian dynamics [237] (and also possibly physics envy [71], p. 61). Fisher's theorem is really about natural selection in the extreme short term [237, 243], and has little to say about the origin of complex traits compared to Darwin's historical method [9–15, 23, 219]. Third, the theory of natural reward suggests that there is no organismal reproductive fitness quantity that natural selection maximizes. Therefore, one can use the simplest measure of organismal reproductive success necessary to solve the problem at hand. Calling this measure *reproductive success* sensu [244], p. 137, rather than *reproductive fitness*, is useful because it avoids connotations of selection. It must be remembered that natural reward can also act through organismal reproductive success.

By abandoning the concept of reproductive fitness, the theory of natural reward also abandons the more particular concept of *inclusive* reproductive fitness (hereafter, *inclusive fitness*). Abandoning the concept of inclusive fitness is useful for avoiding confusion. Many social theorists equate Hamilton's rule with inclusive fitness [245–248]. This allows them to take the empirical support for kinship theory stimulated by Hamilton's rule as support for a larger teleological paradigm [49] based on reproductive fitness maximization [120, 240, 248, 249]. In reality, however, Hamilton's rule is separable from inclusive fitness [250]. Hamilton's rule is a specific statement about the benefit of nepotism [23, 220], which can be derived from any modeling framework [250]. Under the theory of natural reward, Hamilton's rule can be preserved while inclusive fitness and reproductive-fitness-maximization paradigms are abandoned.

What aspects of former theories carry over? To address this, I begin with adaptive landscapes. The adaptive landscape metaphor carries over as long as it is restricted in its application. As conceived by Dobzhansky [31] with peaks representing species, the adaptive landscape can apply to the long-term evolution of simple quantitative traits. Here, peak climbing reflects a trend toward some increased value of a simple trait like body size [32, 71]. However, Dobzhansky's adaptive landscape metaphor does not apply fruitfully to the long-term evolution of very complex traits [71, 251]. To explain the long-term evolution of very complex traits, it is necessary to break the traits into simpler components and apply optimization models to subtraits [1, 9, 23, 214, 219]. Carrying over from Wrightian adaptive landscapes are the ideas that evolutionary advancement involves a trial and error mechanism at a large scale [252], p. 359, and that advancement comes from a process of invention and conquest (dissemination of inventions). The theory of natural reward, however, specifies that invention occurs at the level of species rather than deme, and that natural selection plays a much greater role than drift in assembling inventions (compared to Wright [252] but similar to Wright [200–202]).

With respect to former frameworks of major transitions, two aspects carry over. First, there are many types of evolutionary innovations that have contributed to major ecological expansions, including those involved with individuality [110], cooperation [111, 112], heredity [105], metabolism [253], energy utilization [254], and various other traits [255–257]. Formerly, some authors pitted thematic reviews of evolutionary innovations against each other [105, 110, 112, 258]. However, it will be more useful to ask how some types of innovations synergize, or how understanding some may help illuminate others. A second aspect of major-transition theories carrying over is that devoted to explaining the origin of reproducers [259–261]. Cells, multicellular organisms, and societies all reproduce. The relevant question is how reproduction originates by shifts in life cycles [262–264], and how novel reproducers are maintained as genetically discrete units with fusion-rejection systems [23, 110, 220]. Some former arguments [110, 156, 265] may be recast in terms of transitions in units of reproduction [260].

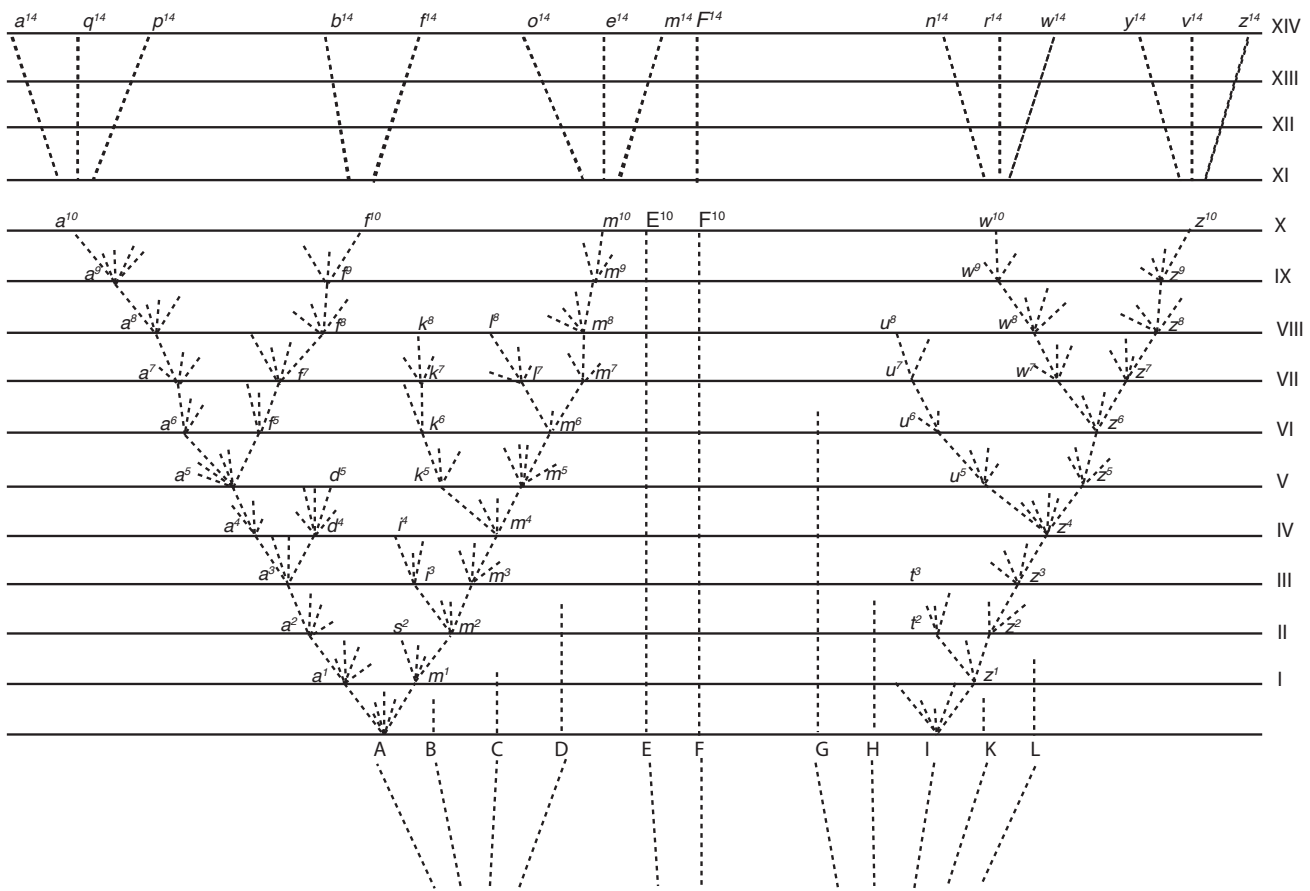

**Figure 8. The long-lost meaning of the only figure in *The Origin*.** The key point of this figure is hidden in plain sight. The initially variable species, A and I, differentially give rise to more descendant species (14) than do the nonvariable species B–H and K–L (1). Darwin hypothesized that variable species would differentially evolve traits allowing seizure of unoccupied or ill-occupied places in the economy of nature, expand and diversify, and avoid extinction. Thus, the differential success of variable species would explain widespread variation in nature, and also suggest long-term progress in evolution toward increased ability to diverge in character (or in modern terms, increased evolvability redrawn from [1]). In editions 1–4, only some of the nonvariable species go extinct. In edition 5, Darwin changed his wording so that all of the nine nonvariable species go extinct (p. 135 of 4th and p. 137 of 5th).

Third, the theory of natural reward will also help repopulate some areas of research by returning them to core Darwinian foundations. The field most likely affected is social theory. As mentioned above, social theorists conflated Hamilton's rule and inclusive fitness, suggesting Hamilton's rule is a sort of general theorem of reproductive fitness. This paradigm led to confusion between cause and effect, and between elementary forms of social behavior [220, 266]. Reorganizing social theory around a Darwinian historical–comparative approach [1, 9, 11, 106], which specifies a narrow role for Hamilton's rule, and which establishes a healthier link between general theory and empirical work [106], will allow for major progress in this field [23, 220].

## Appeal to experts

Several other factors may provide a roadblock for experts. A first factor may be the assumption that innovation requires saltation. This assumption dates back to Darwin, who argued that a finding of graduated steps leading to complex characters disproves a role of a Creator, and thus a role for innovation or "real novelty" in evolution [1], pp. 194, 471; and [267], p. 509. The assumption that innovation requires saltation has led authors advocating novel theories to often unnecessarily advocate saltationism [38, 40]. Recently, authors discussing genetics of macroevolution created an unnecessary dichotomy between their perspectives and classical theories of small incremental changes [77], p. 796, or uniformitarianism [268, 269]. Those defending Darwinism, in turn, have assumed that a ruling out of saltationism disproves the importance of separate analyses of macroevolutionary genetics [36], p. 969; [37] p. 475; [270, 271]; [272], p. 39. Under the theory of natural reward, however, innovation does not require saltation. Inventions can originate gradually and entrepreneurship can still be important for spreading inventions. Thus, ruling out saltation does not disprove natural reward as a force of nature, nor does it suggest that we should ignore the structure of hierarchical gene networks that give insights on macroevolution [77, 270].

Another constraining factor comes from the opinion that if Darwinian competition does not dominate macroevolution, then macroevolution must be governed by chance. This opinion is supported by two dichotomies. The first is between Red Queen (Darwinian) models of direct biotic competition, and Court Jester models of abiotic factors [273]. From the perspective of this dichotomy, a strong role for unpredictable abiotic factors causing climate change or extinction suggests a strong role for chance rather than deterministic forces in macroevolution [273, 274]. This dichotomy ignores, however, that abiotic events can promote indirect competition [61]. The second dichotomy assumes that a finding of a strong role for historical contingency in the origin of complex traits means that it is governed by chance [206, 214, 275]. Under the theory of natural reward, however, a strong role for chance and contingency in the long-term origin of inventions is fully consistent with natural selection acting as nature's tinkerer. Natural reward acts as a non-random entrepreneurial force on the random variation of invention thus provided (Fig. 6).

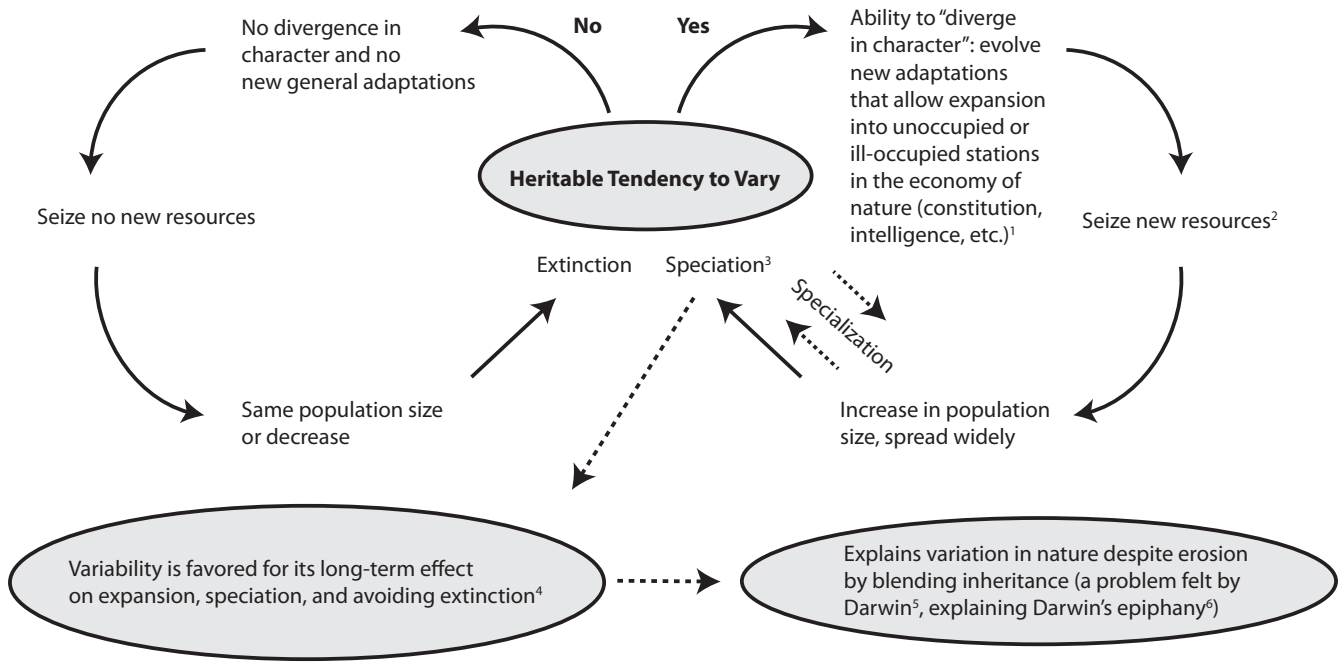

**Figure 9. Meaning of the principle of divergence.** Conceptual foundation of Darwin's principle of divergence. The entire argument is found in [59]. For further details, see: 1. Wallace [278], pp. 105–111. 2. Worster [279], pp. 160. 3. Darwin [1] pp. 104–107. 4. Lyell [4], pp. xiiv-xiv, 327, 459. 5. Fisher [228] 1930, p. 2. 6. Tammone [280], p. 111; Worster [279], p. 160; Gould [38], p. 224.

Evolutionists have also been held back by the opinion, tracing back to Dobzhansky [31], p.77, that Darwin's metaphors like *the struggle for existence* and *survival of the fittest* were not integral to his argument [231]. Many authors have reiterated Dobzhansky's idea that Darwin's metaphors had the unnecessary consequence of suggesting an "eat-or-be-eaten" world. Within this caricature, natural selection is supposed to favor only the most selfish and aggressive individuals through differential mortality of others. Dobzhansky also claimed that such metaphors caused the misuse of Darwinism by propagandists and bigots. Upon this backdrop, evolutionists justify the substitution of the neoDarwinian framework (Fig. 7) for the Darwinian one (Fig. 1). However, if the struggle for existence occurs among alleles, then natural selection acting on genotypes leads to differential net reproductive success of organisms, whether differential fecundity or mortality is the relevant cause. This can cause the survival of altruistic or meek alleles, not just selfish or aggressive alleles. Thus, Darwin's assumptions, when properly used, do not necessarily imply a "eat-or-be-eaten" world. They also do not preclude selection operating through differential fecundity or fertility rather than mortality. Finally, there is no evidence that Darwin's metaphors were responsible for misuse of his theory. Rather, incorrect applications of Darwinism to fields like economics [276] was probably caused by a failure to see the complete picture offered by the theory of natural reward (more discussion of this awaits a future work).

A final question for experts is how the theory of natural reward relates to Darwin's macroevolutionary arguments. I suggest that the theory of natural reward helps reveal the hidden meaning of the only figure in *The Origin of Species* (Fig. 8). Fig. 8 shows two large fans representing variable species, which have a propensity to evolve and adapt, or "diverge in character", expanding and diversifying, and replacing the nonvariable species. Darwin's main argument was that the more variable species would tend to evolve traits allowing seizure of new resources, and thus heritable variability, or tendency to vary [1], p. 118; [59], pp. 252, 255, would be favored by a long-term effect on expansion. Under Darwin's argument, heritable variability allowed adaptation, adaptation allowed population expansion, and population expansion allowed speciation (Fig. 9). Therefore, variability was ultimately favored for its long-term effect on expansion.

Although Darwin's argument for increased variability seems obvious when stated simply, it has not been understood by modern evolutionists. There are four proximate reasons for this. First, Darwin used his figure for three other purposes: to explain the origin of higher taxa through phyletic branching [277], as a model to explain how species diverge in character given variation exists [278, 279], and how specialization can lead to an increase in the number of species on a given resource base [278, 280]. The primary message was lost in the secondary messages. Second, although Darwin argued that the specialization that occurs with speciation feeds back to increase population expansion further (Fig. 9), he also argued that the number of species is ultimately limited by decreasing returns to specialization and ability to evolve by natural selection as populations fragment with speciation [59], pp. 247–248. Most modern evolutionists have not gripped Darwin's entire conceptual scheme, because he did not present the whole thing in *The Origin*. Third, in contrast to modern investigators, Darwin assumed inheritance was blending and searched for factors that would introduce variation back into populations [228], p. 2. For Darwin, the principle of divergence was such a mechanism. Finally, Darwin presented what he perceived to be the evidence for his theory, that larger genera tend to have more variable species than smaller genera, before he presented his theory of divergence itself [1], pp. 54–60, 113–127. This was a very confusing way to present the theory.

Because of their failure to understand his arguments, most neoDarwinists interpreted Darwin's principle of divergence as something more familiar. Their major focus was on forms of speciation (sympatric, allopatric, etc. [203]), so they interpreted the principle of divergence as sympatric speciation [281–284]. From Darwin's perspective, however, the key to understanding the origin of species is not the particular type of speciation, but evolvability and expansion. Darwin's principle of divergence predicts that variability allows the evolution of new advantageous characters, population expansion, and speciation (Fig. 9). In contrast, a lack of variability ultimately leads to population decline and extinction (Fig. 9). Thus, evolvability was key to long-term success. In agreement with Darwin, modern research shows that variation in speciation rates correlate with rates of evolution general morphologies [285], and of various ecologically relevant traits [286], pp. 419–421; [287], but not variation in isolating mechanisms [288]. A shift in focus from types of speciation to innovativeness may yield deeper insights on the origin of species.

Ultimately, Darwin's macroevolutionary theory was obscure because it was incomplete. To explain long-term evolutionary advancement, it is necessary to identify a separate struggle and alternative evolutionary force. Thus when Darwin explained his theory to Lyell [4], pp. xiiv–xiv, 327, 459, Lyell responded by emphasizing what was obvious to many of Darwin's contemporaries [289], pp. 63–64; [290]; [57], pp. 207–208: Darwin's theory raised bigger questions than it answered. Lyell [8] thus referred to Darwin's theory as "one of those humble advances which we may be permitted to make," and cautioned that investigators should, "not be discouraged because greater mysteries remain behind wholly inscrutable to us" [291] pp. 365–366. Alfred Wallace [292], the co-developer of the theory of natural selection, also concluded that, "some more general and more fundamental law underlies that of natural selection..." [292], p. 360.

We now see that these authors were correct. Darwin could not answer the bigger questions because he was ignorant of the form of competition most relevant to macroevolution, the role of mass extinctions, and modern theories of economic growth. He also lived at a time when technological progress was relatively slow compared to today. Darwin thus lacked the language and metaphors necessary to convey the theory of natural reward (Fig. 3 and Table 2). Therefore, just as theories of evolution lacking natural selection could not explain adaptation prior to Darwin [28], neither could Darwin explain progress and advancement without the theory of natural reward.

## Conclusions

Darwin's theory of evolution suggested that some sort of progress and advancement occurred, but it did not explain it. Therefore, a large part of the evolutionary history remained very murky, and this murkiness allowed for two possible interpretations of Darwinism. In the first, natural selection is a short-sighted force that works for the immediate benefits, and only through slight and successive modification. In the second, natural selection is a long-sighted force that works for the ultimate benefits of organismal fitness maximization or selection on long-preserved units (e.g., genes, species, or clades). The theory of natural reward teaches us that only the first interpretation of Darwin's theory is correct. The theory of natural reward also extends Darwinism to explain the apparently progressive trends that must have occurred if Darwin's theory were true. It suggests that the broad-scale history of life is best encapsulated not as *survival of the fittest,* but as *success of the innovative*. The theory of natural reward may also yield important connections between biology and economics not covered here.

## Acknowledgements

This work is the outcome of a long-term research endeavor, the details of which were known only to my closest family and friends until October 2018. The project thus requires unusual acknowledgements. The work was initially motivated by a course taught by ES Vrba and by interactions with LW Buss and his former lab group at Yale University in 1999–2002. I thank JE Strassmann and DC Queller and their former lab group at Rice University for a stimulating intellectual and research atmosphere that led me to key insights in 2008–2011. Special appreciation goes to my family and friends who endured and supported my focused effort over many years. I thank CL Sauer, KA Whitton, JS Gilbert, MC Sledd, CM Bensimon, and TM Gilbert for discussions, general encouragement and support. I thank LE Gilbert for general discussions, helpful comments on a previous version of this manuscript, and long-time advisement. I thank CM Gilbert for support and for proof reading and commenting on the final version of this paper. I appreciate the hospitality of the members of Brackenridge Field Laboratory at the University of Texas at Austin, including RM Plowes, JR Crutchfield, and NT Jones. I thank CV Hawkes and MJ Ryan for sponsoring my position at UT Austin. I thank YE Stuart, JE Strassmann, CE Farrior, JL Martinek and LE Gilbert for commenting on my talk at UT Austin in October 2018.